\documentclass[aps,twocolumn,showpacs,superscriptaddress]{revtex4}
\usepackage{graphicx}
\usepackage{amsmath,amssymb,latexsym}
\newcommand{\WR}{\mathit{Wr}}
\newcommand{\mean}[1]{\langle #1 \rangle}
\newcommand{\meanWR}{\langle |\WR| \rangle}

\begin{document}

\title{Interplay between writhe and knotting for swollen and compact polymers}

\author{Marco Baiesi}
\email[]{baiesi@pd.infn.it}
\affiliation{Instituut voor Theoretische Fysica, K.U.Leuven, Celestijnenlaan 200D, 3001, Belgium.}
\affiliation{Dipartimento di Fisica, Universit\`a di Padova, Via Marzolo 8, 35131 Padova, Italy}

\author{Enzo Orlandini}
\email[]{orlandini@pd.infn.it}
\affiliation{Dipartimento di Fisica, Universit\`a di Padova, Via Marzolo 8, 35131 Padova, Italy}
\affiliation{INFN,  Sezione di Padova, Via Marzolo 8, 35131 Padova, Italy}

\author{Stuart G.  Whittington}
\email[]{swhittin@chem.utoronto.ca}
\affiliation{Department of Chemistry,
University of Toronto, Toronto, Canada M5S 3H6 }


\begin{abstract}
The role of the topology and its relation with the geometry of biopolymers under
different physical conditions
is a nontrivial and interesting problem. 
Aiming at understanding this issue for a related simpler system,
we use Monte Carlo methods to investigate the interplay between writhe and knotting of ring
polymers in good and poor solvents.
The model that we consider is interacting self-avoiding
polygons on the simple cubic lattice.  For polygons with fixed 
knot type we find a writhe distribution 
whose average depends on the knot type but is insensitive to the length $N$ of the polygon
and to solvent conditions. This ``topological contribution'' to the writhe distribution has a value
that is consistent with that of ideal knots. The standard deviation of the writhe increases approximately as
$\sqrt{N}$ in both regimes and this constitutes a
geometrical contribution to the writhe. If the sum
over all knot types is considered, the scaling of the standard deviation changes, for compact
polygons, to $\sim N^{0.6}$. We argue that this difference between the two regimes can be ascribed
to the topological contribution to the writhe that, for compact chains, overwhelms the geometrical
one thanks to the presence of a large population of complex knots at relatively small
values of $N$. For polygons with fixed writhe we find that the knot distribution depends on the
chosen writhe, with the occurrence of achiral knots being considerably suppressed for large writhe.
In general, the occurrence of a given knot thus depends on a nontrivial interplay between writhe,
chain length, and solvent conditions.
\end{abstract}

\maketitle

\section{Introduction}

Important biopolymers such as duplex DNA exist as double-stranded macromolecules, where the
two strands of complementary nucleotides are wound around each other in a right-handed fashion and
around a common axis \cite{WATSON:1953:Nature:13054692}. In addition, the double helix can wind in
space to form a new super-helix, in which case the polymer is said to be supercoiled. A geometric
quantity that has proved to be useful in describing the degree of supercoiling in 
double stranded DNA (dsDNA) is the writhe of a curve, and there is an important conservation theorem
\cite{White:1969,Fuller:1971} relating the writhe of the central
axis of circular dsDNA to the double helical twist and the linking number of the two strands.
Supercoiling in DNA can result from its binding to proteins (histones) in chromatins or from a
linking deficit between the two strands, when the macromolecule becomes a ring, for instance by
cyclization. For example, circular DNA extracted from cells has negative supercoiling
\cite{Bauer:1978:Annu-Rev-Biophys-Bioeng:208457}. On the other hand random cyclization of linear
dsDNA with cohesive ends can trap DNA topoisomers as knots
\cite{Shaw:1993:Science:8475384,Rybenkov:1993:Proc-Natl-Acad-Sci-U-S-A:8506378}. Both supercoiling
and knotting in circular DNA are critical to the functioning of the cell
\cite{Shishido:1987:J-Mol-Biol:2821270} and, for this reason, there exist enzymes to control the
writhe and the knotting of DNA, especially during replication, transcription and recombination.
This suggests that supercoiling and knot formation are relevant indicators of the spatial
arrangement of circular DNA and that a detailed study of their reciprocal influence is crucial in
understanding the conformational properties of this macromolecule at equilibrium
\cite{Podtelezhnikov:1999:Proc-Natl-Acad-Sci-U-S-A:10557257,Burnier:2008:Nucleic-Acids-Res:18658246}.

A relationship between writhe and knot type is evident also in the knot theory of random closed
curves where it is known for example that chiral knots have a non negligible writhe whose sign
depends on the chirality of the knot, and that
there are families of knots (e.g. torus and
twist knots) that can be clearly distinguished in terms 
of their average writhe \cite{BOSTW:1997,Katritch:1996:Nature:8906785}.
For physical knots this relationship is more stringent and allows a classification of knots in
terms of the geometrical (writhe) properties  of their ideal representations
\cite{Katritch:1996:Nature:8906785,ideal-knots:1998,AlessandroFlammini02082007}.

The above mentioned examples refer in general to situations in
which the polymer is in a swollen
phase due to good solvent or unconstrained conditions. There are however several important cases in
which macromolecules such as DNA and proteins are in highly condensed phases due to bad solvent
conditions or to strong confinement. Highly compact configurations have geometrical properties
that clearly differ from their swollen counterparts and this may change the relationship
between writhe and knotting
quite dramatically. This is the case, for example, for highly condensed DNA
extracted from P4 phages \cite{Liu:1981:Nucleic-Acids-Res:6272191,
Wolfson:1985:Nucleic-Acids-Res:3903657,Isaksen:1999:Methods-Mol-Biol:12844863,
Arsuaga:2002:Proc-Natl-Acad-Sci-U-S-A:11959991,Arsuaga:2005:Proc-Natl-Acad-Sci-U-S-A:15958528}. The
genome of the P4 phage is a linear dsDNA, about 11 Kbp long and having two 19bp single stranded
cohesive ends \cite{Murray:1973:Nat-New-Biol:4515740}. In P4 tailless mutants the two ends can move
freely within the capsid and at the end of the packing process they can
cohere to form
rings. Since the genome is very long compared to the linear size of the capsid (50nm)
the DNA is very condensed and a high probability of
knotted molecules is expected. The
experimental data show that over  97\% of the molecules in tailless mutants are indeed knotted.
This is dramatically higher than the 3\% value observed when P4 DNA molecules undergo
cyclization in free solution
\cite{Rybenkov:1993:Proc-Natl-Acad-Sci-U-S-A:8506378,Rybenkov:1997:Science:9235892}. Moreover,
knots in P4 phages are very complex, being characterized by a crossing
number often much
greater than $10$ \cite{Arsuaga:2002:Proc-Natl-Acad-Sci-U-S-A:11959991}. A more detailed analysis
of the extracted knot spectrum based on 2D gel electrophoresis reveals
a predominance of chiral
knots over achiral ones and a prevalence of torus knots over twist knots
\cite{Arsuaga:2005:Proc-Natl-Acad-Sci-U-S-A:15958528,Trigueros:2007:BMC-Biotechnol:18154674}. These
findings suggest a spatial organization of the viral genome characterized by a large amount of
writhe that induces a bias in the observed knot spectrum towards chiral and,
especially, torus knots
\cite{Arsuaga:2005:Proc-Natl-Acad-Sci-U-S-A:15958528,Trigueros:2007:BMC-Biotechnol:18154674}. 
This may suggest an important contribution of the writhe to the condensation of DNA
\cite{Blackstone-et-al-2009}.

All the examples outlined above suggest the importance of a theoretical study of the interplay
between writhe and knot distribution for polymer rings under different equilibrium conditions. Here
we  explore the relation between writhe and knotting
for the two extreme cases of polymers in the swollen and
compact phases.

We perform Monte Carlo simulations on a simple discrete model of a
self-attracting ring polymer,
namely a polygon in the cubic lattice in which the quality of the solvent is
mimicked by assigning an energy gain to any pair of nearest neighbor
non-consecutive vertices of
the polygon. By tuning properly the energy gain this model is known to
describe a collapse
($\Theta$) transition from a swollen phase, characterized by extended configurations, to
one in which the polygons are
compact \cite{Vanderzande,Tesi_ROW96,Tesi_ROW96b}. This
allows us to compare, within the same model, the relationship 
between writhe and knotting in the two regimes
and to see, for example, to what extent the degree of condensation of the polymer affects the
writhe distribution and its relation to the knot spectrum.

The plan of this paper is as follows.
In the next section we describe the model, the Monte Carlo
algorithms and the techniques adopted to compute the writhe and to identify the knot type. 
In Sec.~\ref{sec:res} 
we present the results first for the writhe distribution at fixed knot type and then for the
knot spectrum as a function of the writhe.
Section~\ref{sec:concl} is devoted to a general discussion of the
results presented and to some conclusions.

\section{Model and simulation methods}\label{sec:method}

\subsection{Polymer model}
\begin{figure*}[!tb]
\includegraphics[angle=0,width=7.0cm]{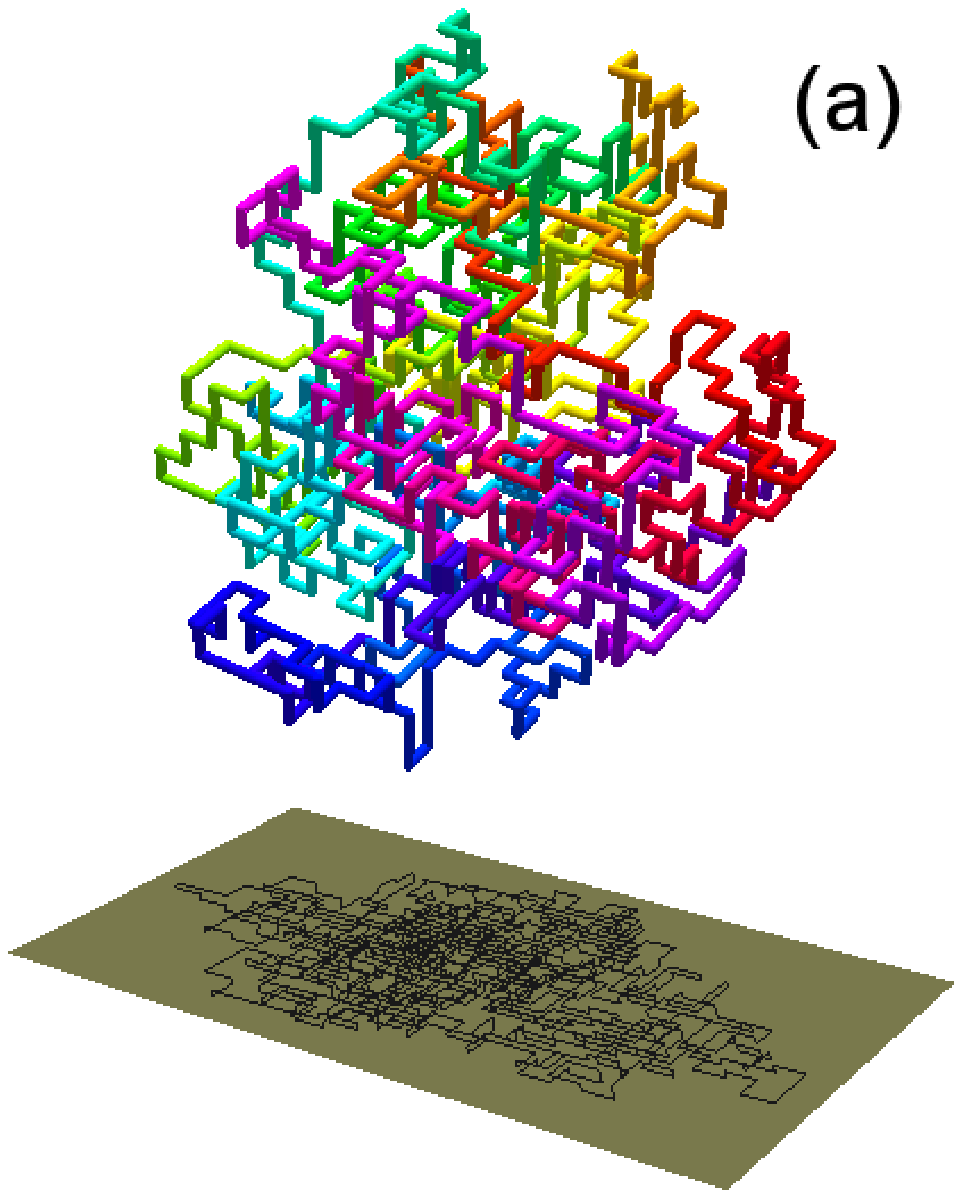}
\hspace{0.3truecm}
\includegraphics[angle=0,width=5.8cm]{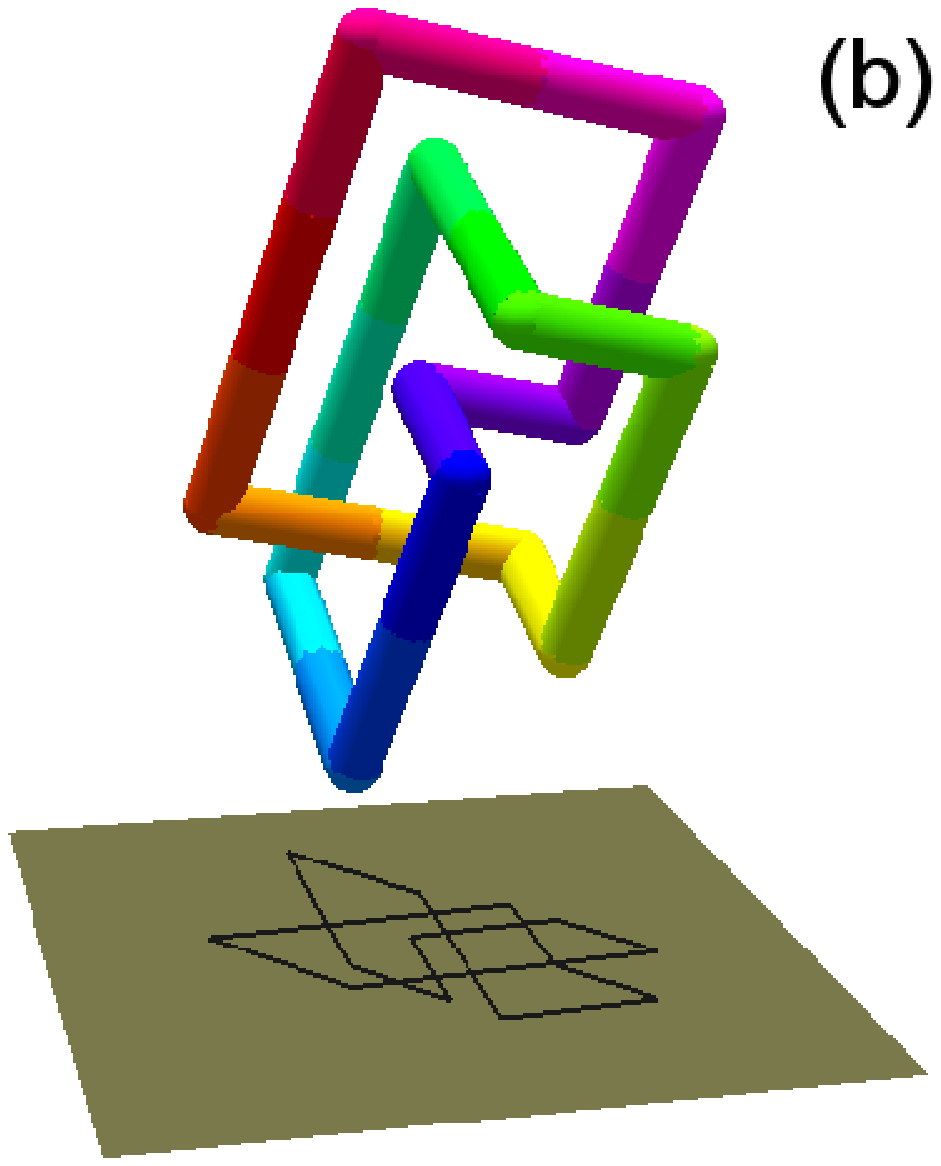}
\caption{(a) Example of a $N=1000$ steps compact configuration and  its projection on a plane. (b)
The configuration of Fig.~\ref{fig:proj}(a) after the simplification procedure based on the BFACF algorithm. Note
that in this case the simplification has been able to reach the minimal number of steps ($N=24$)
compatible with the knot type ($3_1$)
of the polygon \cite{Diao:1993}. \label{fig:proj}}
\end{figure*}

 As a model for the large scale behavior of a long flexible circular polymer in a
good solvent ({\em swollen phase}), we  use $N$-step self avoiding polygons (SAPs) on the
cubic lattice, i.e.~closed lattice walks whose steps can visit each vertex of the
lattice at most once \cite{Vanderzande}. To mimic the quality of the solvent we add to the model an
effective attractive interaction potential which lowers the total energy by $\epsilon =1$ whenever
two non-consecutive vertices of the SAP are one lattice distance apart.
This attractive interaction
is sufficient to induce a collapse transition at a critical temperature $T_c$ and, for $T<T_c$, it
is known to describe the {\em compact phase} \cite{Vanderzande}. In this work we have used
$T=2.5$, well below the value $T_c\simeq 3.717$ \cite{Tesi_ROW96,Tesi_ROW96b}. 
A collapsed configuration with $N=1000$ steps is shown in Fig.~\ref{fig:proj}(a).
Equilibrium
configurations  are sampled by Monte Carlo simulations based on two different algorithms, one for
each equilibrium phase. In the swollen phase the algorithm uses the set of two-pivots moves
i.e.~non-local deformations that are known to sample efficiently unweighted polygons in $Z^3$
\cite{Madras&Slade}.

%%%%%%%%%%%%%%%%%%%%%%%%%%%%%%%%%%%%%%%%%%%%%%%%%%%%%%%%%%%%%%%%%%%
%%%%%%%%%%%%%%%%%%%%%%%%%%%%%%%%%%%%%%%%%%%%%%%%%%%%%%%%%%%%%%%%%%%

In the compact phase the two-pivots algorithm is known to be  inefficient \cite{Tesi_ROW96b}
and in order to have good statistics in this regime we use the pruned enriched Rosenbluth
method (PERM). This is a walk growing algorithm that has been shown to be quite effective in
sampling compact self-avoiding walks~\cite{PERM}. Note that here the efficiency of the algorithm is
mitigated by the fact that only closed chains, i.e.~a small fraction of the whole set of linear
chains, are accepted. Nevertheless this limitation is not very severe in the
compact phase and we have been able to sample compact polygons with $N$ ranging from
$200$ ($680000$ configurations)  up to $1600$ ($220000$ configurations).
For this range of $N$ the compact phase is characterized by a sufficiently high
knotting probability and knot complexity \cite{Baiesi:2007:Phys-Rev-Lett:17930800}. This is not the
case in the swollen phase where the knot population becomes non negligible only for
large values of $N$. In this case, in order to have good statistics of knots, we sample
configurations with $N$ ranging from $1000$ up to $200000$.

\subsection{Computation of the writhe and identification of knots}

A commonly used algorithm to compute the writhe of a curve goes as follows: First, one projects the
curve onto an arbitrary plane. In general the projection will have crossings that most of the time
will be transverse, so that, after having established an orientation of the curve, a sign $+1$ or
$-1$ (determined by a right hand rule)
can be assigned to each crossing. The sum of these signs  gives the signed crossing number
in this projection. The writhe of the curve is obtained by averaging these signed crossing
numbers over all possible projections. From this definition it is clear that the main difficulty in
computing the writhe of a configuration would be the averaging procedure over all projections.
Fortunately, for polygons in $Z^3$ this procedure is enormously simplified by a theorem
\cite{Lacher_Sumners_1991} which reduces the writhe computation to the average of linking numbers
of the given curve with four selected push-offs of the curve itself. In our calculation of the
writhe we made extensive use of this result.

\begin{figure}[!tb]
%\includegraphics[angle=0,width=8.0cm]{Figs/histo_cross_1000_comp.eps}
%\hspace{0.3truecm}
%\includegraphics[angle=0,width=8.0cm]{Figs/histo_cross_1000_swollen.eps}
\includegraphics[angle=0,width=7.5cm]{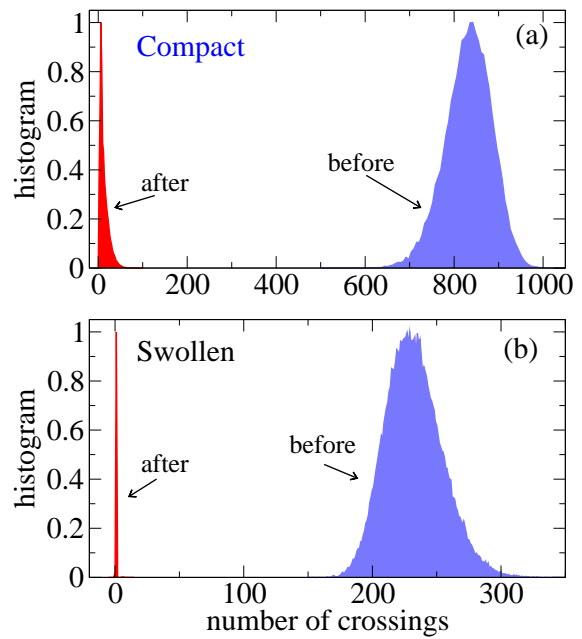}
\caption{Histogram (with arbitrary normalization) 
of the averaged (over 500 projections) number of crossings before
and after the simplification procedure. 
Panels (a) and (b) refer to polygons  with $N=1000$
respectively in the compact and swollen phase. 
 \label{fig:histo_cross_1000}}
\end{figure}

The determination of the knot type of a given configuration, is in general,
a difficult task
both in the swollen phase where $N$ is very large and in the compact phase where
polygons are highly condensed. In both cases any projection of the chain onto a plane gives rise to
a knot diagram with many crossings that makes the calculation of polynomial invariants quite
prohibitive [see Fig.~\ref{fig:proj}(a)]. To circumvent this difficulty, we simplify each sampled
configuration before performing the projection. This is achieved by applying to the polygon a
smoothing algorithm which progressively reduces the length of the chain while keeping its
knot type unaltered (for a similar procedure, 
see \cite{Micheletti:2006:J-Chem-Phys:16483240}). This
procedure is based on the BFACF algorithm \cite{Berg81,Aragao83},
an $N$-varying Monte Carlo method
that is known to be ergodic within each knot type and in which the step fugacity, if kept low
enough, induces a rapid reduction in the number of edges in the
polygon.  This simplification
technique can reduce dramatically the number of crossings encountered in an arbitrary projection,
as shown in Fig.~\ref{fig:proj}(b) for the case of Fig.~\ref{fig:proj}(a) and more generally in
Fig.~\ref{fig:histo_cross_1000} for compact and swollen
polygons with $N=1000$. Note that the simplification procedure
has a dramatic effect on the number of crossings.

%%%%%%%%%%%%%%%%%%%%%%%%%%%%%%%%%%%%%%%%%%%%%%%%%%%%%%%%%%%%%%%%%%%
%%%%%%%%%%%%%%%%%%%%%%%%%%%%%%%%%%%%%%%%%%%%%%%%%%%%%%%%%%%%%%%%%%%

For each simplified configuration we perform $500$ projections and we choose the projection with the
minimal number of crossings. The resulting knot diagram is encoded in terms of the Dowker code
~\cite{Adams}. A further simplification of the Dowker code based on Reidemeister-like moves is
performed. Finally, a factorization of the Dowker code is attempted. This procedure, whenever
successful, splits composite knots into their prime components. From each component of the original
Dowker code we extract, by using {\sc Knotfind}~\cite{knotfind}, the knot type of the original
configuration. In this way we have been able to distinguish
composite knots with up to $5$ prime
components, and with each component having crossing number up to 11.

%%%%%%%%%%%%%%%%%%%%%%%%%%%%%%%%%%%%%%%%%%%%%%%%%%%%%%%%%%%%%%%%%%%
\begin{figure}[!tb]
\includegraphics[angle=0,width=8.0cm]{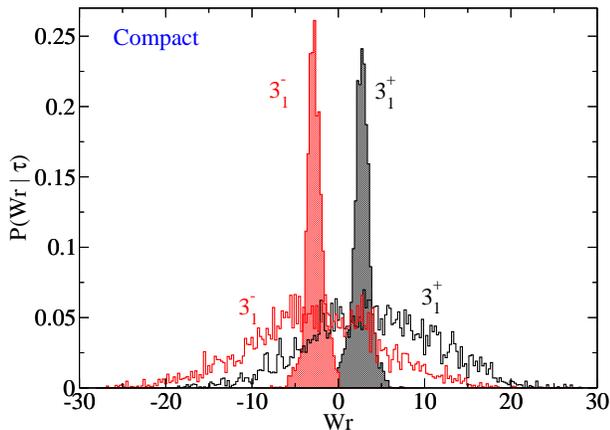}
\caption{The dense histogram is the distribution of the writhe after the simplification based on
the BFACF algorithm for the knot type $3_1$ ($N=1600$, $T=2.5$). This is divided in two sectors
$\WR\le 0$ and $\WR>0$ to emphasize the detection of the two chiralities. Distributions of the
writhe corresponding to original configurations for each chirality are shown as empty histograms.
Bins have size $1/4$, i.e.~the minimum resolution of $\WR$ on the cubic lattice.
\label{fig:Wr-shr}}
\end{figure}
%%%%%%%%%%%%%%%%%%%%%%%%%%%%%%%%%%%%%%%%%%%%%%%%%%%%%%%%%%%%%%%%%%%

To identify the chirality  of a given knot 
type we cannot simply rely on the above procedure, since
the Dowker code does not account for the handedness of the knot. Following
\cite{Micheletti:2008:Biophys-J:18621819} we used a heuristic approach based on the well known
correlation between the chirality of a knot and the writhe of its minimal diagrammatic
representation~\cite{Katritch:1996:Nature:8906785,BOSTW:1997}. This corresponds to computing the
distribution of the writhe of the simplified configurations and looking at the shape of this
distribution. If the distribution is sharply peaked around a well defined positive or negative
value of the writhe one can assert with confidence that the knot is chiral with corresponding sign.
In Fig.~\ref{fig:Wr-shr} we show the distribution of the writhe for the simplified configurations
with a knot $3_1$, for $N=1600$ and $T=2.5$. The distribution is clearly bimodal, and one can
reliably distinguish most of the knot chiralities. Distributions of corresponding configurations
before the BFACF procedure are also shown for comparison. Clearly the main
uncertainties in this
method of chiral detection are due to configurations whose writhe, after BFACF simplification, has
a value close to zero. For these, relatively few, configurations we have calculated the Jones
polynomial, and hence their chirality, explicitly.

\section{Numerical results}\label{sec:res}

%%%%%%%%%%%%%%%%%%%%%%%%%%%%%%%%%%%%%%%%%%%%%%%%%%%%%%%%%%%%%%%%%%%
\begin{figure}[!tb]
\includegraphics[angle=0,width=7.5cm]{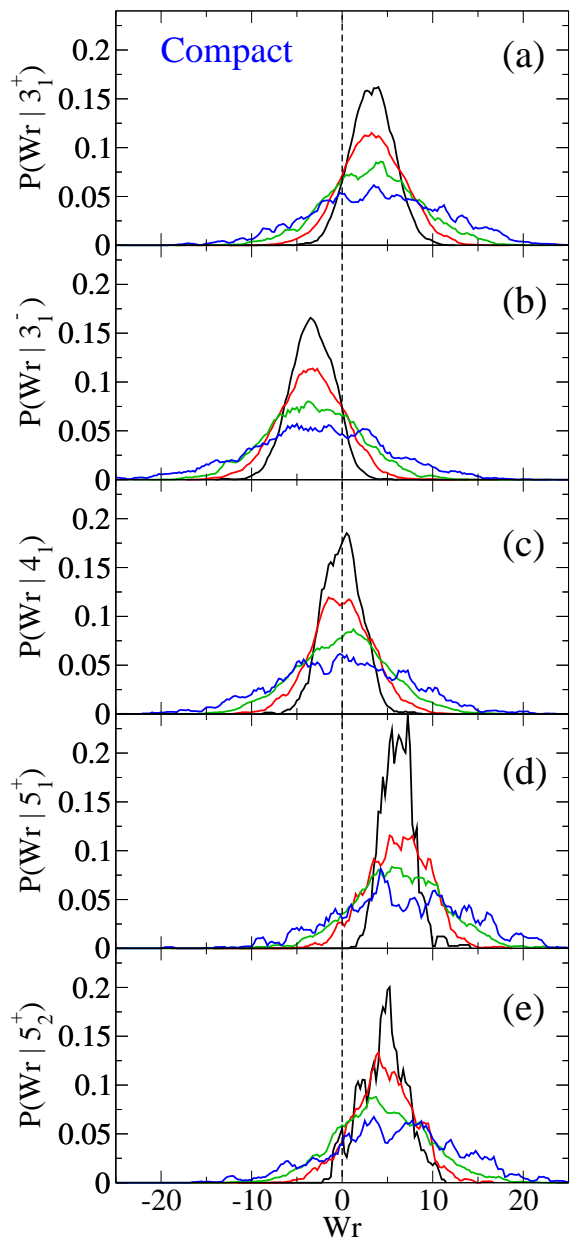}
\caption{ Distributions $P(\WR|\tau)$ in the compact phase for $N=200$ (narrower), $400$, $800$,
and $1600$: (a) $\tau=3_1^+$, (b) $3_1^-$, (c) $4_1$, (d) $5_1^+$, and (e) $5_2^+$.
\label{fig:P(Wr|tau)T2.5}}
\end{figure}
%%%%%%%%%%%%%%%%%%%%%%%%%%%%%%%%%%%%%%%%%%%%%%%%%%%%%%%%%%%%%%%%%%%

%%%%%%%%%%%%%%%%%%%%%%%%%%%%%%%%%%%%%%%%%%%%%%%%%%%%%%%%%%%%%%%%%%%
\begin{figure}[!tb]
\includegraphics[angle=0,width=7.5cm]{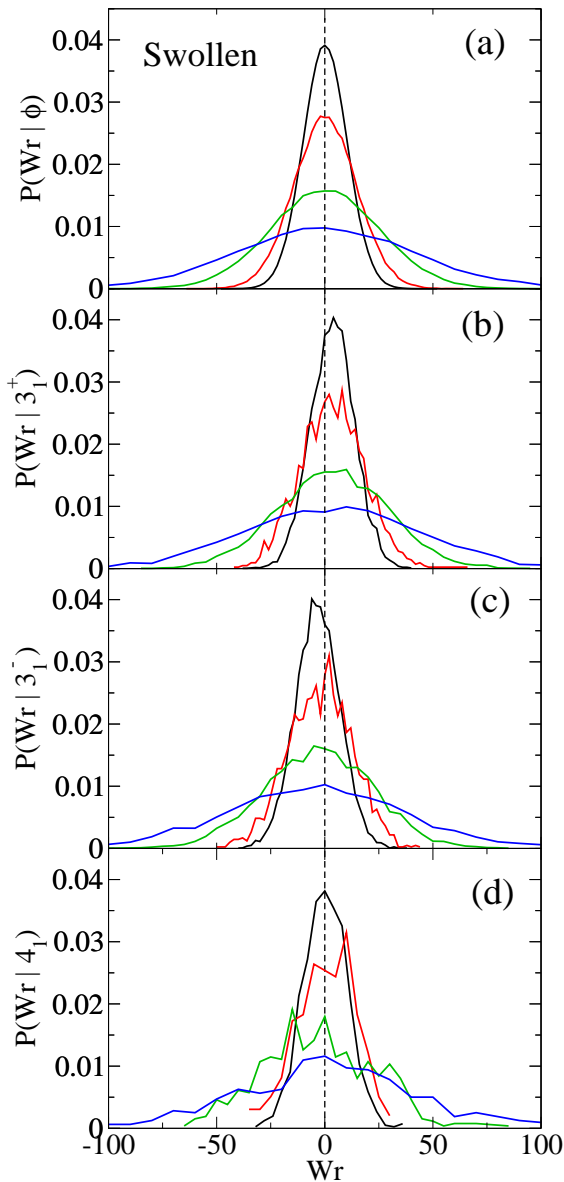}
\caption{ Distributions $P(\WR|\tau)$ in the swollen phase for  $N=5000$ (narrower), $10000$,
$30000$, and $80000$: (a) unknot (b) $\tau=3_1^+$, (c) $3_1^-$, (d) $4_1$. \label{fig:P(Wr|tau)}}
\end{figure}
%%%%%%%%%%%%%%%%%%%%%%%%%%%%%%%%%%%%%%%%%%%%%%%%%%%%%%%%%%%%%%%%%%%

\subsection{The writhe distribution for fixed knot type}
\begin{figure}[!tb]
\includegraphics[angle=0,width=8.0cm]{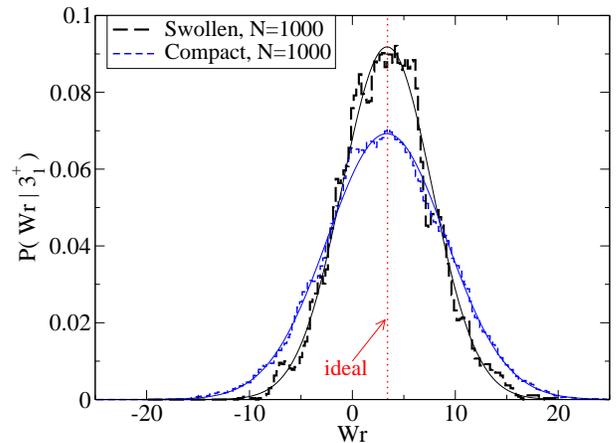}
\caption{ Distributions $P(\WR|\tau)$ for polygons with $N=1000$, both in the swollen (thick dashed
line) and in the compact (dashed line) phase . The thin solid lines correspond to one parameter
Gaussian fits in which the value of the average is fixed to the ideal value $\mu(3_1^+) \simeq
3.41$. \label{fig:P(Wr|tau)N1000}}
\end{figure}

We first focus on the writhe distribution for configurations
with fixed knot type $\tau$,
$P_N(\WR|\tau)$. In Fig.~\ref{fig:P(Wr|tau)T2.5} and Fig.~\ref{fig:P(Wr|tau)} we show  the
distributions $P_N(\WR|\tau)$ for some knots respectively in the compact
and swollen phases. In each
panel different curves correspond to different values of $N$. 
We stress again that, in the good solvent regime, 
the knot probability is very low unless $N$ is large.  Consequently we 
have to consider polygons with $N$ values that are 
up to two orders of magnitude larger
than their compact counterparts. For the $3_1$ knot both
mirror images are shown. The plots suggest a very small (if any)  dependence on $N$ for the average
writhe $\mu(\tau)$ whereas as $N$ increases the distributions tend to broaden. This behavior,
already observed in \cite{BOSTW:1997,Katritch:1996:Nature:8906785} for swollen polygons is here confirmed also for compact
configurations. The different ranges in $\WR$ between swollen and compact phases are due to the
different values of $N$ considered in the two situations. If the same value of $N$ is considered
(see Fig.~\ref{fig:P(Wr|tau)N1000} for $N=1000$) the 
distribution $P_N(\WR|\tau)$ looks
similar in the two phases, being slightly broader for compact polygons.

%%%%%%%%%%%%%%%%%%%%%%%%%%%%%%%%%%%%%%%%%%%%%%%%%%%%%%%%%%%%%%%%%%%
%%%%%%%%%%%%%%%%%%%%%%%%%%%%%%%%%%%%%%%%%%%%%%%%%%%%%%%%%%%%%%%%%%%

%%%%%%%%%%%%%%%%%%%%%%%%%%%%%%%%%%%%%%%%%%%%%%%%%%%%%%%%%%%%%%%%%%%
\begin{figure}[!tb]
\includegraphics[angle=0,width=8.0cm]{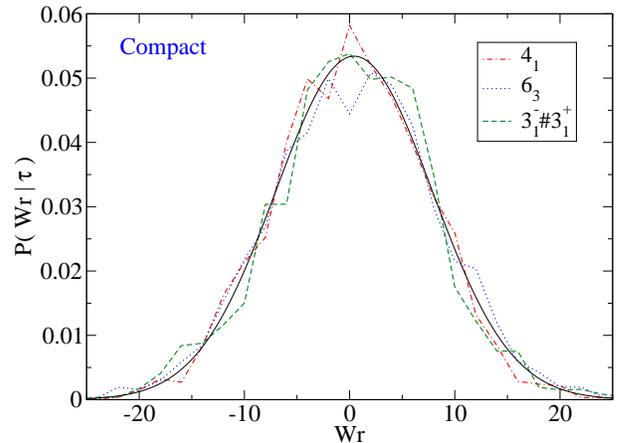}
\caption{ Distributions $P(\WR|\tau)$ for polygons with $N=1600$ in the compact phase, for achiral
knots $\tau=4_1$ and $\tau=6_3$, and for a composite achiral knot $\tau=3_1^-\#3_1^+$. The solid
line is a one parameter Gaussian fit with zero mean.} \label{fig:P(Wr|achiral)}
\end{figure}
%%%%%%%%%%%%%%%%%%%%%%%%%%%%%%%%%%%%%%%%%%%%%%%%%%%%%%%%%%%%%%%%%%%

%%%%%%%%%%%%%%%%%%%%%%%%%%%%%%%%%%%%%%%%%%%%%%%%%%%%%%%%%%%%%%%%%%%
\begin{figure*}[!tb]
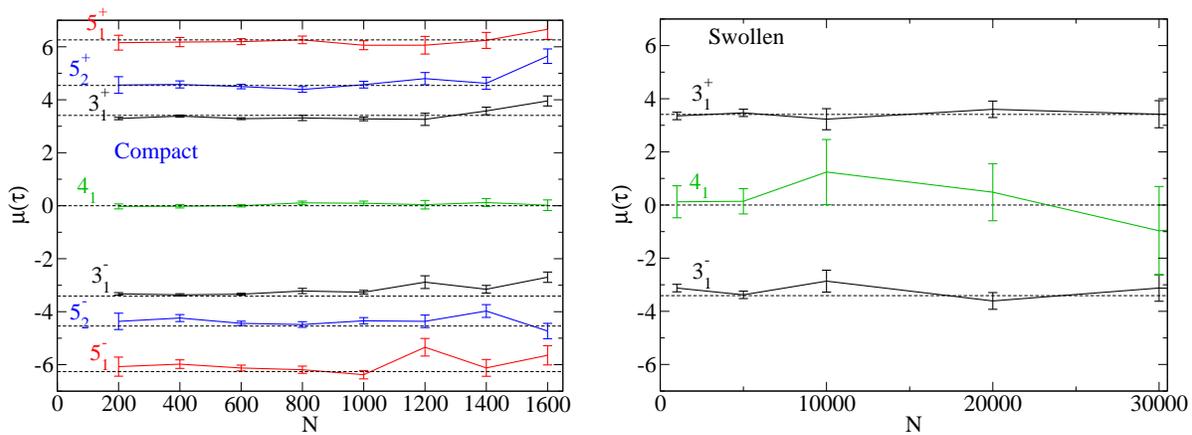

\includegraphics[angle=0,width=7.5cm]{fig_writhe_8_a.eps}
\hspace{0.3truecm}
\includegraphics[angle=0,width=7.7cm]{fig_writhe_8_b.eps}
\caption{ Mean writhe for several knots in the compact (left panel) and swollen (right panel)
phase. Dashed lines are the values of corresponding ideal knots \label{fig:mean}}
%\label{fig:mean:T2.5}}
\end{figure*}
%%%%%%%%%%%%%%%%%%%%%%%%%%%%%%%%%%%%%%%%%%%%%%%%%%%%%%%%%%%%%%%%%%%

%%%%%%%%%%%%%%%%%%%%%%%%%%%%%%%%%%%%%%%%%%%%%%%%%%%%%%%%%%%%%%%%%%%
\begin{figure*}[!tb]
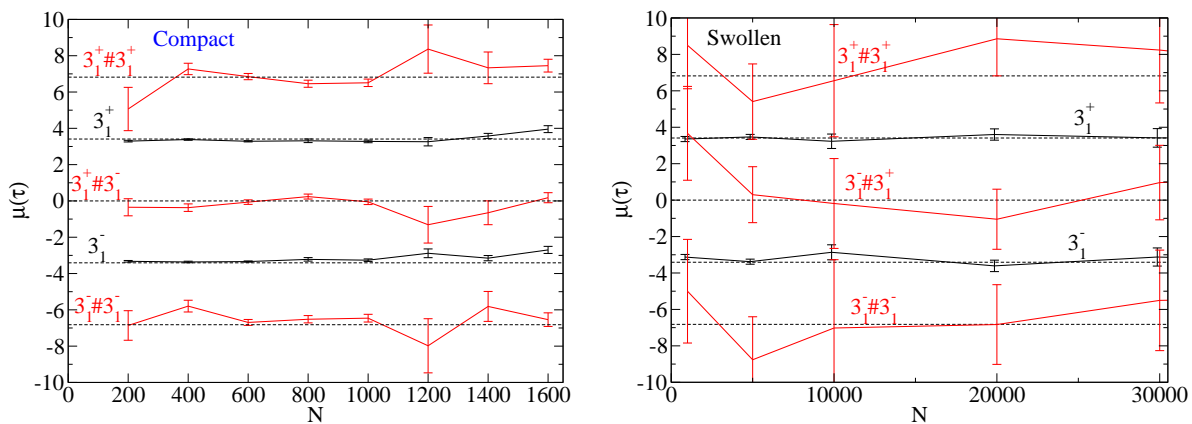

\includegraphics[angle=0,width=7.5cm]{fig_writhe_9_a.eps}
\hspace{0.3truecm}
\includegraphics[angle=0,width=7.7cm]{fig_writhe_9_b.eps}
\caption{ Mean writhe for knots $3_1$ and all the possible composite knots $3_1\#3_1$ in the
compact (left panel) and swollen (right panel) phase. Dashed lines are the values of the sum of the
means writhe of the corresponding ideal prime knots. \label{fig:mean_comp}}
\end{figure*}
%%%%%%%%%%%%%%%%%%%%%%%%%%%%%%%%%%%%%%%%%%%%%%%%%%%%%%%%%%%%%%%%%%%

In Fig. \ref{fig:P(Wr|achiral)} we show the writhe distribution for 
achiral compact polygons for
$N=1600$: all the curves are symmetric and centered around zero. It was shown
rigorously in \cite{BOSTW:1997} that achiral knots have
mean writhe equal to zero for un-weighted polygons.  That argument is a 
symmetry argument that can 
be extended \emph{mutatis mutandis} to 
the case of compact polygons.

In Fig.~\ref{fig:mean} we report the $N$ dependence of the average writhe for different knot types,
respectively in the compact and swollen phase. 
Within error bars ($68\%$ confidence intervals) it
is clear that the average writhe is essentially independent of $N$ and the values are in good
agreement with those calculated previously for different models of polymers in good
solvent~\cite{Podtelezhnikov:1999:Proc-Natl-Acad-Sci-U-S-A:10557257,Burnier:2008:Nucleic-Acids-Res:18658246}
and for ideal knots ~\cite{Katritch:1996:Nature:8906785,ideal-knots:1998} 
(dashed lines in the plots). The agreement between the average 
writhe for a fixed knot type $\tau$  and the value for the corresponding ideal
knot is extended to composite knots as shown in Fig.~\ref{fig:mean_comp} indicating that the
additivity property of the writhe under knot composition found for ideal knots in
~\cite{Katritch:1997:Nature:9217153} and for 
unweighted polygons in~\cite{BOSTW:1997} applies also 
for the case of compact polygons. In fact Sumners \cite{Sumners:2009} has
recently proved a result about additivity of writhe under the 
connect sum operation for unweighted polygons.

The fact that the average writhe is 
independent of the quality of the solvent is a further
indication that this number reflects 
topological properties of the rings no matter how badly they are
embedded in space. One can understand this property by the following argument: for a given
configuration and projection one may attempt to simplify the resulting knot diagram by performing
Reidemeister moves with the goal of removing most of the inessential crossings. This
will give a diagram with the minimal number of crossings compatible with the projection and with
the given knot type $\tau$. Note that in the cubic lattice this procedure would correspond to a
simplification of the original configuration for example by using the BFACF algorithm. This will
bring the original configuration, either swollen or compact, close to the one with the
minimal number of steps (ideal configuration on the cubic lattice). Let us now see how this
procedure affects the average writhe $\mu(\tau)$. Since
the Reidemeister II and Reidemeister III moves always involve pairs
of crossings with opposite signs, the planar writhe 
is unaffected. This is not true for the first
Reidemeister move that removes single 
crossings. For a particular conformation  
the number of crossings with positive and negative signs that
can be removed by Reidemeister I moves will not be 
equal.  However, the procedure of averaging over all 
conformations should drastically reduce this average difference.
The resulting average planar writhe is then related to the
essential crossings for a given knot type and should coincide with the
value for the ideal knot. 
Note that the
above arguments suggest that the 
distribution $P_N(\WR|\tau)$ has Gaussian tails.
In fact if the knot is achiral (as
in Fig.~\ref{fig:P(Wr|achiral)})  or if chiral knots
are split into mirror components (as for example the plus trefoil in
Fig.~\ref{fig:P(Wr|tau)N1000}), the writhe distribution turns out to be
well approximated by a Normal
distribution ${\cal{N}}\left(\mu(\tau),\sigma(\tau)\right)$ 
with mean $\mu(\tau)$ and standard deviation $\sigma(\tau)$
(see solid lines in Fig.~\ref{fig:P(Wr|tau)N1000} and 
Fig.~\ref{fig:P(Wr|achiral)}).

%%%%%%%%%%%%%%%%%%%%%%%%%%%%%%%%%%%%%%%%%%%%%%%%%%%%%%%%%%%%%%%%%%%
\begin{figure*}[!tb]
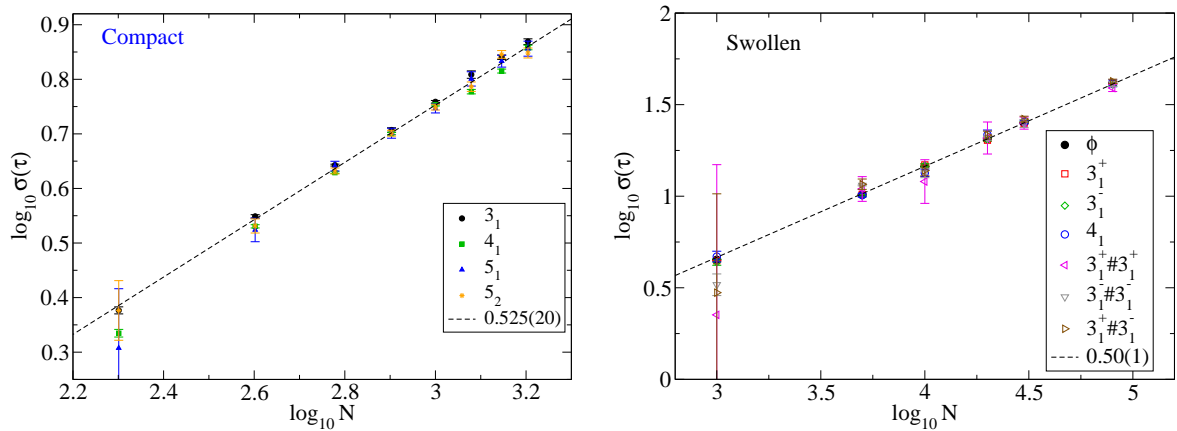

\includegraphics[angle=0,width=7.5cm]{fig_writhe_10_a.eps}
\hspace{0.3truecm}
\includegraphics[angle=0,width=7.5cm]{fig_writhe_10_b.eps}
\caption{Log-log plot of the standard deviation $\sigma_N(\tau)$ 
of the  writhe distribution, as a function of $N$, for several knots
(averaged over mirror images), in the compact (left panel) and swollen (right panel)
phase. For each knot type the data have been fit by a power law of the form $A(\tau)
N^{\eta_{\tau}}$ (dashed lines) with $\eta_{\tau}=0.525\pm 0.020$ in the compact phase and
$\eta_{\tau}=0.50\pm 0.01$ in the swollen one. Note that for compact polygons, in order to avoid
correction to scaling effects, the fit has been obtained  by excluding the data $N=200,400$.
\label{fig:variance_compact_swollen}}
\end{figure*}
%%%%%%%%%%%%%%%%%%%%%%%%%%%%%%%%%%%%%%%%%%%%%%%%%%%%%%%%%%%%%%%%%%%

By looking at the width of the distributions in Figures~\ref{fig:P(Wr|tau)T2.5}  and
\ref{fig:P(Wr|tau)} it is clear that the standard deviation $\sigma_N(\tau)$ increases with $N$ and this dependence is
shown in Fig.~\ref{fig:variance_compact_swollen}, left panel, for compact prime knots and for
swollen knots in the right panel. A simple fit of the data in the form $A(\tau)
N^{\eta_{\tau}}$ gives estimates of the exponents that, within error bars, are independent on the
knot type $\tau$ and whose value is $0.525 \pm 0.020$ for compact polygons and $0.50 \pm 0.01$
for swollen ones. Note that the two values are, within error bars, compatible suggesting that
$\eta_{\tau}$ is also independent of the equilibrium phase, and possibly equal to $1/2$.

%%%%%%%%%%%%%%%%%%%%%%%%%%%%%%%%%%%%%%%%%%%%%%%%%%%%%%%%%%%%%%%%%%%
\begin{figure*}[!tb]
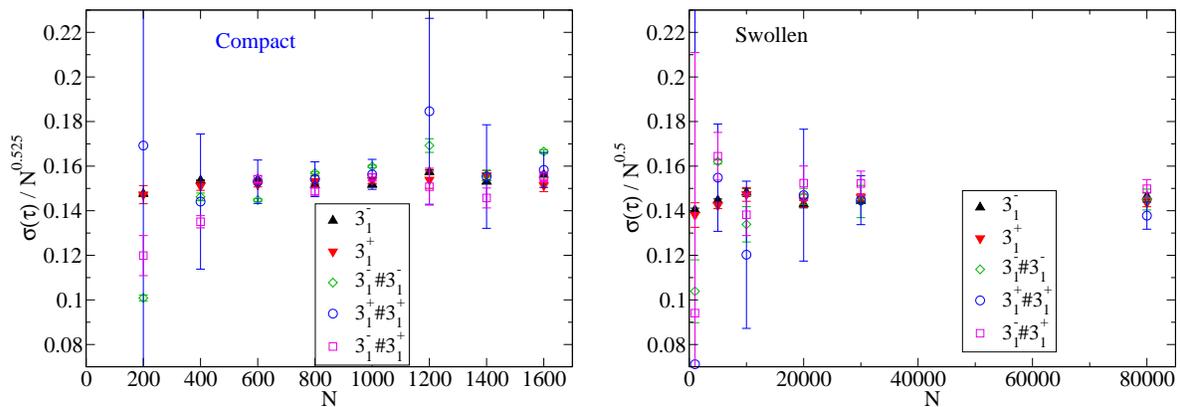

\includegraphics[angle=0,width=7.5cm]{fig_writhe_11_a.eps}
\hspace{0.3truecm}
\includegraphics[angle=0,width=7.5cm]{fig_writhe_11_b.eps}
\caption{Standard deviation $\sigma(\tau)$ divided by $N^{\eta_{\tau}}$ for different knot types in
the compact (left panel) and swollen (right panel) phase. The value of $\eta_{\tau}$ considered
is  $0.525 \pm 0.020$ for compact polygons and $0.50 \pm 0.01$ for the swollen ones.
\label{fig:var_Wr_div_N0.5_31_3131}}
\end{figure*}
%%%%%%%%%%%%%%%%%%%%%%%%%%%%%%%%%%%%%%%%%%%%%%%%%%%%%%%%%%%%%%%%%%%

Given that $\eta_{\tau}$ seems to be independent of $\tau$ it is interesting to see if this is also
the case for the amplitude $A(\tau)$. In Fig.~\ref{fig:var_Wr_div_N0.5_31_3131}  we plot
$\sigma_N(\tau)/ N^{\eta_{\tau}}$ for different knot types, respectively in the compact (left
panel) and swollen (right panel) phase. Within error bars the amplitude turns out to be independent
of the knot type.

\begin{figure}[!tb]
\includegraphics[angle=0,width=8.0cm]{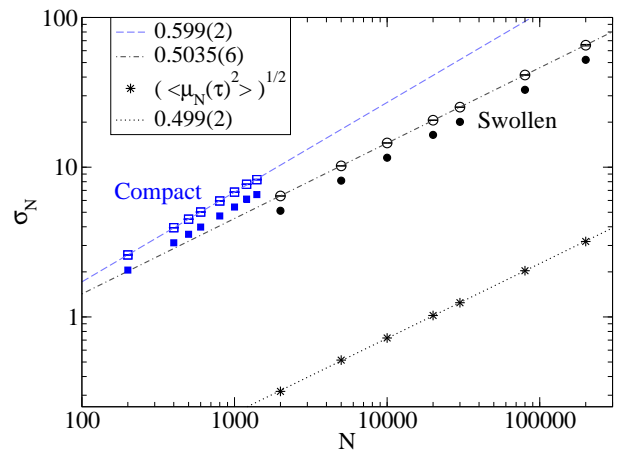}
\caption{Log-log plot of $\sigma(\tau)$ vs $N$ in the swollen phase (circles) and in the compact
phase (squares). The filled symbols refer to the mean of the absolute value of the writhe. Dashed
lines are the result of a linear fit and have a slope of $0.599\pm 0.002$ and $0.5035\pm 0.0006$
respectively for compact and swollen polygons.
Stars refer to the average square of the mean value of writhe of knots in the swollen phase
and the dotted line is a fit with slope $0.499\pm 0.002$.
\label{fig:variance_N}}
\end{figure}

It is interesting to compare the $N$ dependence of $\sigma_N(\tau)$ [i.e.~the standard deviation of
the distributions $P_N(\WR|\tau)$] with the one, $\sigma_N$, that measures the broadness of
$P_N(\WR)=\sum_{\tau} \pi_N(\tau)P_N(\WR|\tau)$ where $\pi_N(\tau)$ is the probability of occurrence of
knot type $\tau$. Note that $P_N(\WR)$ is the writhe distributions of the whole set of polygons. In
Fig.~\ref{fig:variance_N} we plot $\sigma_N$ as a function of $N$ for swollen (empty
circles) and compact (empty squares) polygons. Moreover, since often in
the literature, the width of
the writhe distribution $P_N(\WR)$,  is described in terms  of the mean of the absolute value of
the writhe $\meanWR_N$, in Fig.~\ref{fig:variance_N} we have also reported this quantity in terms
of  solid symbols. The plots are on a log-log scale and the linear behavior suggests a power law
dependence $A N^{\eta}$. Note that for each phase solid and empty symbols lie on parallel lines
indicating that $\sigma_N$ and $\meanWR_N$ display the same power law behavior but different
amplitude. A simple linear fit of the data furnishes the estimate $\eta =0.5035\pm 0.0006$ in the
swollen phase and $\eta = 0.599 \pm 0.002$ in the compact one. The value found for the swollen
regime agrees with previous estimates \cite{BOSTW:1993} and is close to the rigorous lower bound
$1/2$ established for $\meanWR$ \cite{BOSTW:1993}. In the compact phase, however, the exponent
$\eta$ is significantly higher than its swollen counterpart but markedly smaller than the rigorous
upper bound $4/3$ found in \cite{Cantarella:2002}. Its value is also smaller than $0.75$, i.e.~the
one estimated in \cite{ Micheletti:2006:J-Chem-Phys:16483240}, for Gaussian chains confined in
small spheres. The difference between these two last estimates could be due to the presence, in our
case, of the excluded volume interactions that increase the persistence length of the
chain making the wrapping of the chain around itself more costly.  However,
the difference in the value of $\eta$ could be due to a 
difference in the spatial organization
of compact configurations obtained by self-attraction 
compared to the ones obtained by a severe geometrical 
confinement.

An important byproduct of the analysis of $\sigma_N$ and $\sigma_N(\tau)$ is that
$\eta \approx \eta_{\tau}$ in the swollen phase whereas $\eta > \eta_{\tau}$ 
in the compact one. This should be contrasted
with the fact that $\eta_{\tau}$ is nearly the same $\simeq 1/2$ in the two phases.

A possible explanation of this striking behavior may be that the 
writhe of a given configuration can be decomposed into a ``geometrical" contribution coming from the
spatial organization of the polygon and the topological one coming purely from 
the knot type of the configuration. 
For the swollen case it is known that the 
geometrical contribution to $\sigma$ increases at least as 
rapidly as $\sqrt N$ which is what was found for the set of 
all polygons in the swollen phase.  
In this phase, the topological contribution 
apparently does not scale faster than 
$\sqrt{N}$. 
In general however the situation could be different since a sum over all
knot types must be taken into account and the topological contribution could become important if
the probability of a knot type $\tau$ to occur, i.e.~$\pi_N(\tau)$ is 
sufficiently large. This turns out
to be the case for compact polygons where the knotting probability increases with $N$ with an
exponential factor that is a factor of $1000$ bigger than for 
its swollen counterparts. 

To make these arguments
more precise let us suppose that $P_N(\WR|\tau)$ are well approximated by Gaussian distributions. If
this is the case (as suggested by our results) the
conditional distributions 
$P_N(\WR|\tau)$ are completely defined  by the average
$\mu_N(\tau)$ and the standard deviation $\sigma_N(\tau)$. 
If knots are distinguished by chirality,
clearly  the variance of the distribution 
is $\sigma_N(\tau)^2$. If we sum over the mirror images the mean 
writhe is zero and the overlap of
two Gaussian distributions implies 
that $\mean{\WR^2}_N(\tau) = \sigma_N(\tau)^2+\mu_N(\tau)^2$. This term
is for fixed knot type and, in order to compute  
the global second moment, the term must be weighted
by the probability of occurrence of the knot
\[
\mean{\WR^2} = \sum_\tau \pi_N(\tau) \mean{\WR^2}_N(\tau).
\]
This gives
\[
\mean{\WR^2} = \mean{\sigma_N(\tau)^2} +\mean{\mu_N(\tau)^2}.
\]
The scaling of $\langle Wr^2\rangle$ depends on whether $\langle
\sigma^2\rangle$ or $\langle \mu^2\rangle$ increases more rapidly
as $N$ increases.  For the compact case it seems that 
$\langle \mu^2\rangle$ increases more rapidly than
$\langle \sigma^2\rangle$ while this is not true in the swollen phase.
We can estimate $\mean{\mu_N(\tau)^2}$ in the swollen phase by 
using the ideal $\mu_N(\tau)$ whenever a knot $\tau$ contributes to the
statistics.
Since the two members of a chiral pair occur with roughly 
equal probabilities and since knots are weakly localized in the 
swollen phase we can estimate the writhe of a composite knot 
(for all chiral combinations) by assuming independence.
The average $(\mean{\mu_N(\tau)^2})^{1/2}$ from this analysis turns out to scale $\sim\sqrt{N}$,
see Fig.~\ref{fig:variance_N}, and with an amplitude much smaller than $\sigma_N$,
confirming that the scaling of $\sigma_N$ is governed by the geometrical
spread $\mean{\sigma_N(\tau)^2}$ of all knots around their ideal value.

%%%%%%%%%%%%%%%%%%%%%%%%%%%%%%%%%%%%%%%%%%%%%%%%%%%%%%%%%%%%%%%%%%%
%%%%%%%%%%%%%%%%%%%%%%%%%%%%%%%%%%%%%%%%%%%%%%%%%%%%%%%%%%%%%%%%%%%

Hence, the important difference between the  swollen phase and the compact one is that collapsed
polymers display a rich spectrum of knots $\pi_N(\tau)$ for moderate lengths
$N$~\cite{Baiesi:2007:Phys-Rev-Lett:17930800}, while in good solvent conditions
knots start to appear with a non-negligible
frequency only for  $N\gtrsim 10^5$ \cite{Orlandini:2007:Rev-Mod_Phys}. At that stage
$\sigma(\tau)\gg |\mu(\tau)|$ and hence $\mean{\WR^2}_N(\tau) \simeq \sigma(\tau)^2$. On the
contrary, in the compact phase knots start to appear with a non-negligible frequency already for
$N$'s where $|\mu(\tau)|\gtrsim \sigma(\tau)$, in which case  $\mean{\WR^2}_N(\tau) \simeq
\mu(\tau)^2$ is a good approximation. Because of this effect, the global $\mean{\WR^2}$ has a
scaling that does not trivially follow that of single knots.

\subsection{Knot spectrum at fixed writhe}
\begin{figure}[!tb]
\includegraphics[angle=0,width=8.0cm]{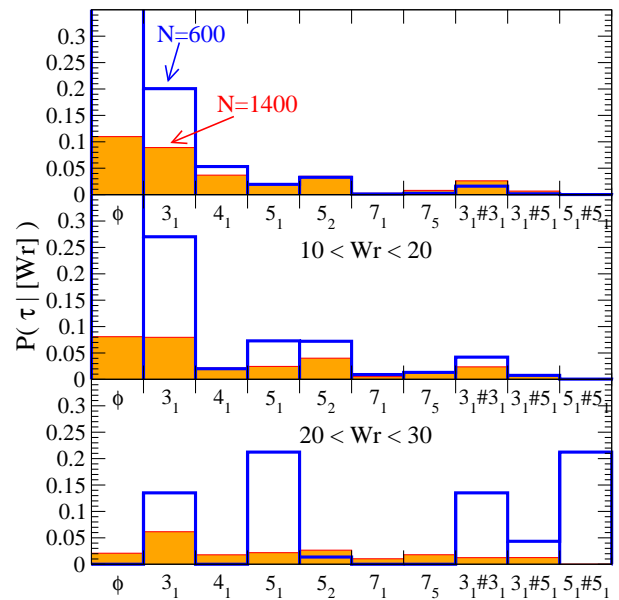}
\caption{
Frequency of some knots, regardless of the writhe (upper panel), for compact configurations with
$10<\WR<20$ (central panel), and  $20<\WR<30$ (lower panel), both for $N=600$ (empty histogram)
and for $N=1400$ (dense histogram).
\label{fig:Wr_windows_2}}
\end{figure}

Podtelezhnikov {\em et al}~\cite{Podtelezhnikov:1999:Proc-Natl-Acad-Sci-U-S-A:10557257}
have investigated the probability of finding a specific knot type,
conditioned on the writhe, for a model of circular DNA in a good solvent.
In this section we examine how the presence of writhe
might affect the knot spectrum in
the swollen and compact phases.  The idea is to extract a subset
of the sample of polygons whose writhe falls into a window of
values $[\WR_{\min},\WR_{\max}]$. Within this subset one looks
at the relative frequency of occurrence of a given knot. This
corresponds to estimating the
probability
$P(\tau | {[\WR_{\min},\WR_{\max}]})
= \int_{\WR_{\min}}^{\WR_{\max}}P(\tau|\WR) /
\sum_\tau\int_{\WR_{\min}}^{\WR_{\max}}P(\tau|\WR) $. In Fig.~\ref{fig:Wr_windows_2} we plot the
knot spectrum for two different intervals of the writhe (middle
and lower panels) for polygons in the
compact phase. The top panel is the knot spectrum obtained by summing over all values of the
writhe and is shown as a reference. Let us focus first on the $N=600$ case (empty bars). Here it is
clear that, as writhe increases, the knot spectrum changes dramatically with respect to the case of
unconstrained writhe (top panel). In particular one sees that in configurations with
relatively large value of writhe (bottom panel) the achiral knot $4_1$ is practically suppressed
and that the torus knot $5_1$ becomes more frequent than the twist knot $5_2$. This is an effect
that was previously found for almost ideal configurations
\cite{Podtelezhnikov:1999:Proc-Natl-Acad-Sci-U-S-A:10557257,Burnier:2008:Nucleic-Acids-Res:18658246} 
and for polymers confined in spheres
\cite{Arsuaga:2005:Proc-Natl-Acad-Sci-U-S-A:15958528}. Here
it is extended to the case of random
self-attracting polymers in the collapsed phase.

In previous sections we have shown that the average writhe of a given knot
is insensitive to changes in $N$.
It thus makes sense to use the same windows $[\WR_{\min},\WR_{\max}]$ for any comparison
between results for different $N$'s. On the other hand, we have seen that the variance
$\sigma_N(\tau)^2 \sim N$ indicating that writhe distributions for contiguous (in complexity) knots
will have an overlap that becomes more and more important as $N$ increases. Hence, a windowing
procedure should be less efficient in picking up a given subset of knots. This is indeed the case,
for compact polygons with $N=1400$ (filled histograms in Fig.~\ref{fig:Wr_windows_2}). For
$N=1400$ the values of the writhe considered do not show the writhe bias effect on the knot
spectrum because the writhe distributions of the knots considered are not well separated.

%%%%%%%%%%%%%%%%%%%%%%%%%%%%%%%%%%%%%%%%%%%%%%%%%%%%%%%%%%%%%%%%%%%
%%%%%%%%%%%%%%%%%%%%%%%%%%%%%%%%%%%%%%%%%%%%%%%%%%%%%%%%%%%%%%%%%%%

In the swollen phase, the conditional distributions $P(\tau|\WR)$ of different knots can be of
comparable magnitude in a window of values $[\WR_{\min},\WR_{\max}]$ only for very large $N$,
because knots are rare for short chains. For long chains,  $P(\tau|\WR)$ is essentially a broad
Gaussian with average much smaller than its standard deviation, and of course with normalization
$\sim P(\tau)$. Hence, in a window $[\WR_{\min},\WR_{\max}]$ one just normally finds that $ P(\tau|
[\WR_{\min},\WR_{\max}]) \sim  P(\tau)$, that is, no sensible dependence on the writhe interval is
present.

\section{Conclusions}\label{sec:concl}
We have investigated with Monte Carlo methods 
the interplay between writhe and knotting
in ring polymers in both good and poor solvent conditions.  To
model good solvent conditions we used self-avoiding polygons on
the simple cubic lattice and we incorporated an attractive
interaction between neighboring pairs of vertices 
to model poor solvent conditions, leading to compact configurations.

For polygons with fixed knot type we computed the writhe distribution.
The mean writhe is zero for achiral knots while for chiral knots
it is non-zero but is insensitive to the length of the polygon.  This
has been observed previously for the good solvent case but we observe
the same effect for poor solvents.  
Moreover, for each knot type, both mean values are consistent with that
of ideal knots.

In the swollen case,
the width of the writhe distribution increases as the square root of the
length of the chain.
Within errors, this is true also in the collapsed phase.
On the other hand, the mean spread of the writhe  computed without distinguishing knots
scales differently (with exponent $\simeq 0.6$) for compact polymers.
The reason is that, in this regime, the population of knots grows quickly with the chain length,
including complex knots with large mean writhe values.
Therefore, the swollen phase is characterized by the geometrical spread of configurations, 
while the statistics of writhe in the compact regime is governed by 
the topological contribution of the mean writhe of knots.

Finally, we have observed that picking up a specific knot by constraining the writhe in a 
given window is not possible if the chain length is much longer than
the persistence length of the polymer.

It is well known that the mobility of a circular DNA molecule in
gel electrophoresis is sensitive to the writhe of the molecule.
If the circular DNA is prepared with a particular
knot type then this type of experiment could be used, in principle,
to measure the mean writhe and determine how the writhe is
related to the knot type.  By changing the ionic strength
of the solution the typical dimensions of the molecule
could be changed, in an analogous way to changing solvent quality.

\section{Acknowledgment} 
E.O. was supported by a grant from MIUR-PRIN05.
S.G.W.~acknowledges financial support from NSERC.
 M.B. acknowledges financial support from K.~U.~Leuven under
      Grant No. OT/07/034A and from University of Padua 
      and Progetto di Ateneo n. CPDA083702.

\end{document}